\documentclass[aps,twocolumn]{revtex4}
\usepackage{amssymb}
\usepackage{amsmath}
\usepackage{graphicx}
\usepackage{epsfig}
\newtheorem{theorem}{Theorem}
\newtheorem{lemma}{Lemma}

\begin{document}

\title{Friedmann Robertson Walker models with Conformally Coupled Massive
Scalar Fields are Non-integrable}
\author{L. A. A. Coelho}
\email{l_coelho2@yahoo.com.br}
\affiliation{Programa de P\'os-Gradua\c{c}\~ao em F\'isica, Universidade do Estado do Rio de Janeiro, Rua S\~ao Francisco Xavier 524, Maracan\~a, Rio de Janeiro - RJ,
20550-900, Brazil}

\author{J. E. F. Skea}
\email{jimsk@dft.if.uerj.br}
\affiliation{Departamento de F\'isica Te\'orica, Instituto de F\'isica, Universidade do Estado do Rio de Janeiro, Rua S\~ao Francisco Xavier 524, Maracan\~a, Rio de Janeiro - RJ,
20550-900, Brazil}

\author{T. J. Stuchi}
\email{tstuchi@if.ufrj.br}
\affiliation{Instituto de F\'isica, Universidade Federal do Rio de Janeiro, Ilha do Fund\~ao, Caixa Postal 68528, Rio de Janeiro - RJ, 21945-970, Brazil}

\begin{abstract}

In this work we use a recently developed nonintegrability theorem
of Morales and Ramis to prove that the Friedmann Robertson Walker
cosmological model with a conformally coupled massive scalar field is nonintegrable.

\end{abstract}

\maketitle

\section{Introduction}

\ \indent  In recent years the search for nonintegrablility criteria for Hamiltonian systems
 in the complex domain has acquired more relevance~\cite{mora}-\cite{yosh}. 
Such techniques are potentially of particular importance
in cosmology because of controversies over both integrability, and the existence of chaos
in cosmological models~\cite{tere}-\cite{bianchi}. Part of the problem is that certain
methods used to traditionally measure chaos in non-relativistic systems, such as the Lyapunov
exponents, are no longer valid in General Relativity where there is no absolute time
coordinate.

  In this work we use a recently developed theorem by Morales and Ramis~\cite{mora} which
establishes a relation between two differents concepts of integrability: the complete
integrability of complex analytical Hamiltonian systems (given by Liouville's theorem) and the
integrability of homogeneous linear ordinary differential equations (LODEs) in terms of
Liouvillian functions in the complex plane. A Liouvillian function is a function
which can be written as a combination of elementary functions, algebraic functions
(solutions of polynomial equations), their indefinite integrals or exponentials of these integrals. Since we are working in the complex domain
this definition includes elementary functions such as logarithms, trigonometric functions
and their inverses.

  The model we study in this paper is a Friedman-Robertson-Walker (FRW) cosmological model with a conformally coupled
scalar field, $\phi$, of mass $m$. We use the conformal form of the metric
\begin{equation}
{\rm d}s^2=a^2(\eta)\left[{\rm d}\eta^2-\frac{1}{1-kr^2}\,{\rm d}r^2-r^2\,{\rm d}\Omega^2\right],
\label{line}
\end{equation}
with $\eta$ the conformal time, $a(\eta)$ the scale factor and $k=0,\pm 1$ the curvature.

  The dynamics of this model has been discussed and studied before using numerical
methods~\cite{tere}-\cite{hamils} but, as far as we are aware, no completely
rigorous conclusion has been reached about its integrability. In~\cite{hami2} the integrability of a generalisation of the
model studied here is considered using Painlev\'e analysis via the ARS
algorithm~\cite{ARS1,ARS2,ARS3}. Though there is a strong connection between integrability and the Painlev\'e property, and the latter has been remarkably successful in indicating possibly integral cases, it is worth noting that the lack of the Painlev\'e property is not a rigorous obstruction to
integrability~\cite{Grammaticos}. Additionally, the ARS algorithm is not a foolproof method for determining whether a system possesses the Painlev\'e property, and its application
can lead to false conclusions~\cite{mix1,mix2,mix3}, particularly when applied to determining the non-integrability of a dynamical system. Finally, certain expressions in~\cite{hami2}
are undefined for our model, requiring a separate analysis.

  The Morales-Ramis Theorem (MRT) which we use in our study rigorously provides necessary conditions for the integrability of a Hamiltonian system and so sufficient conditions for non-integrability.
The theorem can be used to reduce the question of integrability to one of the existence
of Liouvillian solutions of a homogeneous second-order linear ODE. This problem can in
turn be solved using Kovacic's algorithm. Though complex to write down,
the algorithm is, as we shall see, straightforward to apply to the problem
considered here. 
To use the MRT for our cosmological model we first require a
Hamiltonian which generates the
field equations. In this case it is known~\cite{tere}-\cite{hamils} that a suitable
Hamiltonian is

\begin{equation}
 H=\frac{1}{2}\left[(p_\phi^2+k\phi^2)-(p_a^2+ka^2)+m^2a^2\phi^2\right]=0,
 \label{Hamil}
\end{equation}

\noindent where $p_a$ and $p_\phi$ are the momenta conjugate to $a$ and $\phi$ respectively.

In the next section we give the main results of the Morales-Ramis theorem.
Since the various versions of Kovacic's algorithm in the literature~\cite{mora,bianchi,kova,duval}
have slight differences in presentation and conventions, we include the version of the
algorithm as used by us. We then show how the
algorithm quickly determines that the Hamiltonian is nonintegrable for $k\neq 0$. Finally
for the case $k= 0$ the analysis based on the invariant planes $a=p_a=0$ and $\phi=p_\phi=0$
is inconclusive. However, because the potential is homogeneous in this case, there exist
particular nonsingular solutions which do not lie in these planes which can be used as
a basis for the analysis. Fortunately the case of homogeneous potentials has been
exhaustively studied by Yoshida~\cite{yosh} and Morales-Ramis~\cite{mora} and so we
can simply apply those results.

\section{The Morales-Ramis Theorem}

 The Morales-Ramis theorem is a nonitegrability criterion: it gives
a necessary condition for a Hamiltonian system to be integrable and therefore
a sufficient condition for nonintegrability. The
theorem is based on the analysis of the variational equations
(in particular the normal variational equation, or NVE) for the
perturbations of a non-equilibrum particular solution. The basic idea is that if the flow of
the Hamiltonian system has a regular behaviour (is integrable), then
the linearized flow along a particular integral curve given by the
NVE must also be regular (integrable). Conversely
if the linearized flow is nonintegrable the system as a whole will
be nonintegrable.

  A Hamiltonian system, $X_H$, of dimension $n$ is called integrable if there
exist $n$ independent constants of the motion in involution. By considering the
differential Galois group of the NVE, the theorem
of Morales-Ramis links this concept of integrability to an apparently
different concept of integrability -- the existence of Liouvillian
solutions of the NVE of $X_H$. The theorem may be stated as

\begin{theorem}
If there are $n$ first integrals of $X_H$ that are independent and in involution,
then the identity component of the Galois group of the NVE is abelian.
\end{theorem}

It is known that~\cite{kapla} for an ODE to admit a Liouvillian solution, the identity component of its Galois group must be soluble. Hence, if the solutions are not Liouvillian, the identity component of the Galois group is not soluble and, therefore, non-Abelian.

Our strategy will therefore be:

\noindent{\bf 1:} Select a particular solution (in our case an invariant plane).\\
\noindent{\bf 2:} Write the variational equations and the NVE.\\
\noindent{\bf 3:} Check if the solutions of the NVE are Liouvillian functions.\par

To decide the third step, we use Kovacic's algorithm~\cite{kova} which
we now turn to describe.


\section{Kovacic's algorithm}

\ \indent Kovacic's algorithm provides a procedure for
computing the Liouvillian solutions of a homogeneous linear second order differential 
equation. If the algorithm terminates negatively, we can
conclude that no such solutions exist.

Let ${\mathbb{C}}(x)$ be the field of rational complex functions
(ratios of polynomials in $x$ with
complex coefficients). It is well-known that by using the change of dependent variable
\begin{equation}
 y=\xi\,\exp\left({1\over 2}\int b\,{\rm d}x\right)
\label{trans1}
\end{equation}
the second order homogeneous LODE
\[
y''+b(x)\,y'+c(x)\,y = 0
\]
can be transformed to the so-called reduced invariant form
\begin{equation}
\xi''-g\xi=0,
\label{kova1}
\end{equation}
where
\begin{equation}
g(x)=\frac{1}{2}{b}'(x)+\frac{1}{4}b(x)^2-c(x).
\label{gdef}
\end{equation}
Note that, if
$b(x)$ and $c(x)~\in{\mathbb{C}}(x)$ then $g(x)~\in{\mathbb{C}}(x)$. 

Moreover, using a further change of variables $v=\xi'/\xi$, equation~(\ref{kova1})
is transformed into the Riccati equation 
\begin{equation}   
v'+v^2=g.
\label{kova2}
\end{equation}

Now equation~(\ref{kova1}) is integrable, if and only 
if equation~(\ref{kova2}) has an algebraic solution, that is $v$ solves a 
polynomial equation $f(v)=0$, where the degree of $f$ (the minimal polynomial) 
in $v$ belongs to the set $L=\left\{1,2,4,6,12\right\}$.

Kovacic's algorithm can be divided into three main steps:
the first step is the determination of the subset of $L$ relevant for the LODE
under consideration; the two other steps
are devoted respectively to determining the existence of the 
minimal polynomial, and its construction. If the algorithm does not terminate
successfully (ie, equation~(\ref{kova2}) has no
algebraic solution) then equation~(\ref{kova1}) has no solution in 
terms of Liouvillian functions.

In the version used of the algorithm we essentially follow~\cite{mora,bianchi,duval,luis}. Let
\begin{equation}
g=g(x)=\frac{s(x)}{t(x)},
\label{gx}
\end{equation}
with $s(x),t(x)$ relatively prime polynomials, and $t(x)$ 
monic. Define the function $h$ on the set 
$L_{max}=\{1,2,4,6,12\}$ by $h(1)=1$, $h(2)=4$, $h(4)=h(6)=h(12)=12$.\par

{\bf Step 1} (determination of possible orders of the minimal polynomial)

If $t(x)=1$ then set $m=0$, else factorize $t(x)$ into monic relatively 
prime polynomials 
\[t(x)=t_1(x)\,t^2_2(x)\ldots t^m_m(x),\]
\noindent where $t_i$ have no multiple roots and $t_m\neq1$.

Then\par

{\bf 1.1} Let $\Gamma'$ be the set of roots of $t(x)$ (i.e.,the 
singular points in the finite complex plane) and let $\Gamma=\Gamma'\cup 
\infty $ be the set of singular points.

 Then the order of a singular point 
$c\in \Gamma'$ is, as usual, $o(c)=i$ if $c$ is a root of multiplicity 
$i$ of $t_i$. The order at infinity is defined by 
$o(\infty)=\mbox{max}(0,4+\mbox{deg}(s)-\mbox{deg}(t))$. We call 
$m^+=\mbox{max}(m,o(\infty))$. 

 For $0\leqslant i\leqslant m^+$, denote by 
$\Gamma_i=\left\{c\in\Gamma\mid o(c)=i\right\}$ the subset of all elements of order $i$.

{\bf 1.2} If $m^+\geq 2$ then we write 
$\gamma_2=\mbox{card}(\Gamma_2)$, else $\gamma_2=0$. Then we compute\par

\centerline{$\displaystyle{\gamma=\gamma_2+\mbox{card}\left( 
\bigcup_{\stackrel{\mbox{\scriptsize$3\leq k\leq m^+$}}
                   {k~\mbox{odd}}
               }         \Gamma_k\right)}.$}

{\bf 1.3} For the singular points of order one or two, $c\in \Gamma_2 
\cup \Gamma_1$, we compute the principal parts of $g$:\par

\centerline{$\displaystyle{g_c=\alpha_{c}(x-c)^{-2}+\beta_{c}(x-c)^{-1}+O(1),}$}\par

\noindent if $c\in \Gamma'$, and\par

\centerline{$\displaystyle{g_{\infty}=\alpha_{\infty}x^{-2}+\beta_{\infty}x^{-3}+O(x^{-4}),}$}\par

\noindent for the point at infinity.\par

{\bf 1.4} We define the subset $L'$ (of all possible values for the 
degree of minimal polynomial) as $\{1\}\subset L'$ if $\gamma=\gamma_2$, 
$\{2\}\subset L'$ if $\gamma \geq 2$ and $\{4,6,12\}\subset L'$ if $m^+ 
\leq 2$.\par

{\bf 1.5} We have the three following mutually exclusive cases:\par

     1.5.1  If $m^+ > 2$, then $L=L'$.

     1.5.2  Define $\Delta_c = \sqrt{1+4\alpha_c}$. If $m^+ \leq 2$ and $\forall c\in \Gamma_1\cup\Gamma_2$, $\Delta_c\in {\mathbb Q}$, then $L=L'$.\par

     1.5.3  If cases (1.5.1) and (1.5.2) do not hold, then 
$L=L'-\{4,6,12\}$.

{\bf 1.6} If $L={\emptyset}$, then equation~(\ref{kova1}) is 
non-integrable with Galois group $SL(2,{\mathbb{C}})$, else one writes $n$ for the 
minimum value in $L$.\par


For the second and third steps of the algorithm we 
consider a fixed value of $n$.\par

{\bf Step 2}\par

{\bf 2.1} If $\infty$ has order $0$ we write the set\par

\centerline{$\displaystyle{E_{\infty}=\left\{0,\frac{h(n)}{n},2\frac{h(n)}{n},3\frac{h(n)}{n},\ldots,n\frac{h(n)}{n}\right\}.}$}\par

{\bf 2.2} If $c$ has order 1, then $E_{c}=\{h(n)\}$.\par

{\bf 2.3} If $n=1$, for each $c$ of order 2 we define\par

\centerline{$\displaystyle{E_{c}=\left\{\frac{1}{2}(1+\Delta_c),\frac{1}{2}(1-\Delta_c)\right\}}$}\par

{\bf 2.4} If $n\geq 2$, for each $c$ of order 2, we define\par
\begin{equation*}
E_{c}={\mathbb Z}\cap 
\left\{\frac{h(n)}{2}(1-\Delta_c)+\frac{h(n)}{n}k\Delta_c: 
k=0,1,\ldots ,n\right\}.
\end{equation*}

{\bf 2.5} If $n=1$, for each singular point of even order $2\nu$, with 
$\nu >1$, we compute the numbers $\alpha_c$ and $\beta_c$ defined (up 
to a sign) by the following conditions:\par

\begin{quote}
     2.5.1 If $c\in \Gamma'$,\par
\end{quote} 


\centerline{$\displaystyle{
g_c=\left\{{\alpha_{c}\over (x-c)^{\nu}}
           +\sum_{i=2}^{\nu-1}{\mu_{i,c}\over (x-c)^{i}}\right\}^2
           +{\beta_c\over\left(x-c\right)^{\nu+1}}+
O\left(x-c\right)^{-\nu}
}$}

\noindent and we write \par

\centerline{$\displaystyle{\sqrt{g_c}:=\alpha_c\left(x-c\right)^{-\nu}+\sum_{i=2}^{\nu-1}{\mu_{i,c}\left(x-c\right)^{-i}}.}$}\par

\begin{quote}
     2.5.2  If $c=\infty $,\par
\end{quote} 

\centerline{$\displaystyle{g_{\infty}=\left\{\alpha_{\infty}x^{\nu-2}+\sum_{i=0}^{\nu-3}{\mu_{i,\infty}x^{i}}\right\}^2-\beta_{\infty}x^{\nu-3}+O\left(x^{\nu-4}\right),}$}\par

\noindent and we write\par

\centerline{$\displaystyle{\sqrt{g_{\infty}}:=\alpha_{\infty}x^{\nu-2}+\sum_{i=0}^{\nu-3}{\mu_{i,\infty}x^{i}}.}$}\par

Then for each $c$ as above, we compute\par

\centerline{$\displaystyle{E_c=\left\{\frac{1}{2}\left(\nu+\epsilon 
\frac{\beta_c}{\alpha_c}\right):\epsilon =\pm 1\right\},}$}\par

\noindent and the sign function on $E_c$ is defined by\par

\centerline{$\displaystyle{\mbox{sign}\left(\frac{1}{2}\left(\nu+\epsilon 
\frac{\beta_c}{\alpha_c}\right)\right)=\epsilon,}$}\par

\noindent being $+1$ if $\beta_c=0$.\par

{\bf 2.6} If $n=2$, for each $c$ of order $\nu$, with $\nu \geq 3$, we 
write $E_c=\{\nu\}$.

{\bf Step 3}\par

{\bf 3.1} For $n$ fixed, we try to obtain elements ${\bf e}=(e_c)_{c\in 
\Gamma}$ in the Cartesian product $\prod_{c\in \Gamma}^{}{}E_c$, such 
that:\par

{(i) $\displaystyle{d({\bf e}):= n-\frac{n}{h(n)}\sum_{c\in 
\Gamma}^{}{}e_c}$ is a non-negative integer,}

(ii) If $n=2$ or $n=6$ then {\bf e} has an even number of elements
which are odd integers

(iii) when $n=4$, then {\bf e} has at least two elements not
divisible by 3, and the sum of all elements not divisible by 3 is divisible by 3.

If no such set ${\bf e}$ is obtained, we select the next value in $L$ 
and repeat Step~2, else $n$ is the maximum value in $L$ and 
the Galois group is  $SL(2,{\mathbb{C}})$ (and equation~(\ref{kova1}) is non-integrable).\par

{\bf 3.2} For each family ${\bf e}$ as above, we try to obtain a 
rational function $Q$ and a polynomial $P$, such that\par

(i) \centerline{$\displaystyle{Q=\frac{n}{h(n)}\sum_{c\in 
\Gamma'}^{}{}\frac{e_c}{x-c}+\delta_{n1}\sum_{c\in \bigcup_{\nu 
>1}^{}{}\Gamma_{2\nu}}^{}{}\mbox{sign}(e_c)\sqrt{g_c},}$}

\noindent where $\delta_{n1}$ is the Kronecker delta.\par

(ii) $P$ is a polynomial of degree $d({\bf e})$ and its coefficients 
are found as a solution of the (in general, overdetermined) system of 
equations\par

\centerline{$\displaystyle{P_{-1}=0,}$}\par
\centerline{$\displaystyle{P_{i-1}=-(P_i)'-QP_i-(n-i)(i+1)gP_{i+1},~~n\geq 
i\geq 0,}$}\par
\centerline{$\displaystyle{P_n=-P.}$}\par

If a pair $(P,Q)$ as above is found, then equation (\ref{kova1}) is 
integrable and the Riccati equation (\ref{kova2}) has an algebraic 
solution $v$ given by any root $v$ of the equation\par

\centerline{$\displaystyle{f(v)=\sum_{i=0}^{n}{\frac{P_i}{(n-i)!}v^i=0.}}$}\par

If no pair as above is found we take the next value in $L$ and we go to 
Step~2. If $n$ is the greatest value in $L$ then the Galois
group of~(\ref{kova1}) is $SL(2,{\bf{C}})$ and the ODE is non-integrable.\par

\section{Application and Result}

\subsection{The case $k\neq 0$}

We apply the theorem of Morales-Ramis to~(\ref{Hamil}). We choose as
our set of non-equilibrium particular solutions the invariant plane $p_a=a=0$. 
The NVEs relative to this plane are
\begin{equation}
\frac{{\rm d}^2\delta a}{{\rm d}t^2}=\left(-k+m^2\phi^2\right)\delta a.
\label{kova4}
\end{equation}
 
Changing the independent variable to $\phi$ and renaming
$\delta a=y$, we obtain the equation

\begin{equation}
\frac{{\rm d}^2 y}{{\rm d}\phi^2}+\frac{1}{\phi}\frac{{\rm d}y}{{\rm d}\phi}
+\left(\frac{m^2}{k} -\frac{1}{\phi^2}\right)y=0.
\label{eq45} 
\end{equation}

This equation is a second-order, linear and homogeneous ODE
with coefficients which are rational functions of $\phi$, and we can
therefore apply Kovacic's algorithm to determine any Liouvillian solutions.

Using~(\ref{trans1}) and~(\ref{gdef}) we transform~(\ref{eq45}) into the reduced
invariant form~(\ref{kova1})

\begin{equation}
\xi''= \left(\frac{3k-4m^2\phi^2}{4k\phi^2}\right)\,\xi.
\label{kova7}
\end{equation}

\begin{lemma}
Equation~(\ref{eq45}) has no Liouvillian solutions when $k\neq 0$ and $m\neq 0$.
\end{lemma}

{\it Proof}: by application of Kovacic's algorithm to equation~(\ref{kova7}).

{\bf Step 1}

\ \indent $g(\phi)$ has one finite pole, at $\phi=0$, of order 2 and the pole at infinity, of order 4
 (since, by assumption, we are treating the massive case, $m\neq 0$).
 This implies that $m^+=4$ and $\gamma=\gamma_2=1$. Since the pole at $\phi=0$ belongs to $\Gamma_2$,
 we calculate the Laurent series (when $k\neq0$) as

\[g_0=\frac{3}{4}\phi^{-2}-\frac{m^2}{k}\]

\noindent Hence $\displaystyle{\alpha_0=\frac{3}{4}}$ and $\beta_0=0$. Thus we have $L=\{1\}$.

{\bf Step 2}

\ \indent Because $L=\{1\}$ the unique value for $n$ is $n=1$. Through the items $2.3$ and $2.5$
 we calculate the sets $E_c$. From $2.3$ we have that
 $\displaystyle{E_0=\left\{\frac{3}{2},-\frac{1}{2}\right\}}$.
 In item~$2.5.2$ we need to expand $g$ around $\phi=\infty$. Doing this we obtain

\[g_\infty=\frac{3}{4}\phi^{-2}-\frac{m^2}{k}\Longrightarrow E_\infty=\{1\}.\]

Summarizing, 

\[E_0=\left\{\frac{3}{2},-\frac{1}{2}\right\} ~~~~~~\mbox{and}~~~~~~E_\infty=\{1\} .\] \par

{\bf Step 3}

\ \indent In this step we need to calculate $\prod_{c\in \Gamma} E_c$, using
 the sets determined in the previous step. We obtain the set of sets given by

\[\prod_{c\in \Gamma}^{}{}E_c= \left\{ \left\{\frac{3}{2},1\right\},\left\{-\frac{1}{2},1\right\} \right\}.\]

From 3.1(i) we calculate the values of $d(\bf{e})$ as $\displaystyle{d=-\frac{3}{2}}$ and $\displaystyle{d=\frac{1}{2}}$
 respectively. Since neither of theses values satisfies 3.1(i) and there are no other values of $n$ in $L$,
 the Galois group of~(\ref{kova7}) is $SL(2,{\mathbb{C}})$, equation (\ref{kova7}) is nonintegrable in terms of
 Liouvillian functions, and therefore the system represented by the Hamiltonian~(\ref{Hamil}) is also nonintegrable when $k\neq0$. This completes the proof.

\subsection{The case $k=0$}

\begin{lemma}
When $k=0$ the only first integral of the Hamiltonian system~(\ref{Hamil}) is the
Hamiltonian, and the system is therefore nonintegrable.
\end{lemma}

{\it Proof:} In order to prove this lemma we observe that when $k=0$ the Hamiltonian
(\ref{Hamil}) has a homogeneous potential. Hamiltonians with homogeneous potentials have
been exhaustively studied using the MRT and particular results obtained which we now
outline.

Let

\begin{equation}
 H={1\over 2}\sum_{i=1}^n p_i^2 + V(q_1,\ldots,q_n)
\label{hamhomog}
\end{equation}

\noindent where $A$ is a constant and $V$ is a homogeneous potential,
i.e. $V(A\overrightarrow{Q})=A^{g}\,V(\overrightarrow{Q})$ with $g$ being the degree of the potential. To put our Hamiltonian in the form~(\ref{hamhomog}) we perform the canonical transformation, $x=ia$ and $P_x=-iP_a$, after which

\begin{equation}
 H=\frac{1}{2}\left[p_x^2+p_\phi^2-m^2x^2\phi^2\right]
\label{hamhom}
\end{equation}

\noindent and $g=4$. The MRT for homogeneous potentials is given by~\cite{mora,yosh}.

\begin{theorem}
Let $V(q_1, \ldots, q_n)$ be a homogeneous potential function of integer degree
$g$, $c$ a solution of the equation $c=\overrightarrow{V}'(c)$, and $\lambda_i$
(the Yoshida coefficients) the eigenvalues of the matriz $V''(c)$. One of these
eigenvalues is trivial, in that it corresponds to the tangential variational equation,
and has value $g-1$.

If a Hamiltonian system of the form~(\ref{hamhomog}) is completely integrable (with holomorphic or meromorphic first integrals) then each pair $(g,\lambda_i)$ belongs to
one of the following list (where we do not consider the trivial case $g=0$)
\begin{equation}
\begin{array}{ll}
(1) & (g,p+p(p-1)g/2)                               \\ 
(2) & (2,\mbox{arbitrary complex number})                  \\ 
(3) & (-2,\mbox{arbitrary complex number})                  \\ 
(4) & (-5,\frac{49}{40}-\frac{1}{40}(\frac{10}{3}+10p)^2)     \\ 
(5) & (-5,\frac{49}{40}-\frac{1}{40}(4+10p)^2)          \\ 
(6) & (-4,\frac{9}{8}-\frac{1}{8}(\frac{4}{3}+4p)^2)     \\ 
(7) & (-3,\frac{25}{24}-\frac{1}{24}(2+6p)^2)             \\ 
(8) & (-3,\frac{25}{24}-\frac{1}{24}(\frac{3}{2}+6p)^2)    \\ 
(9) & (-3,\frac{25}{24}-\frac{1}{24}(\frac{6}{5}+6p)^2)     \\ 
(10) & (-3,\frac{25}{24}-\frac{1}{24}(\frac{12}{5}+6p)^2)    \\ 
(11) & (3,-\frac{1}{24}+\frac{1}{24}(2+6p)^2)                 \\ 
(12) & (3,-\frac{1}{24}+\frac{1}{24}(\frac{3}{2}+6p)^2)        \\ 
(13) & (3,-\frac{1}{24}+\frac{1}{24}(\frac{6}{5}+6p)^2)         \\ 
(14) & (3,-\frac{1}{24}+\frac{1}{24}(\frac{12}{5}+6p)^2)         \\ 
(15) & (4,-\frac{1}{8}+\frac{1}{8}(\frac{4}{3}+4p)^2)             \\ 
(16) & (5,-\frac{9}{40}+\frac{1}{40}(\frac{10}{3}+10p)^2)          \\ 
(17) & (5,-\frac{9}{40}+\frac{1}{40}(4+10p)^2)                      \\ 
(18) & (g,\frac{1}{2}(\frac{g-1}{g}+p(p+1)g))                        \\ 
\end{array}
\end{equation}

\noindent where $p$ is an arbitrary integer.
\end{theorem}

For the system represented by~(\ref{hamhom})
the only non-trivial Yoshida coefficient is $\lambda=-1$.
For $g=4$ the only possibilities for satisfying Theorem~3 are (1), (15) and (18).
For all these cases there are no integer values of $p$ which solve $\lambda=-1$, and
we conclude that the system represented by~(\ref{hamhom}) is nonintegrable.
This completes the proof of lemma~2.

We can now enunciate the following theorem

\begin{theorem}
 The Friedmann Robertson Walker model with a conformally coupled massive scalar field represented by the Hamiltonian~(\ref{Hamil}) is not completely integrable.
\end{theorem}

{\it Proof:} By lemma~1 there are no Liouvillian solutions of the NVE for the plane $a=p_a=0$. This implies that the identity component of its Galois group is not soluble and therefore non-Abelian.  Using theorem~1 we have that the only first integral of the Hamiltonian system is the Hamiltonian itself, and that the system is not completely integrable in this case. By lemma~2 the Hamiltonian system is also not completely integrable when $k=0$. Therefore the Hamiltonian system represented by~(\ref{Hamil}) is nonintegrable for all values of $k$.

\section{Conclusion}

\ \indent

From our analysis we have shown rigorously using analytic methods that
FRW universes with a conformally coupled massive scalar field are nonintegrable.
This is compatible with results from numerical analysis based on
Poincar\'e sections~\cite{tere} which indicate that the behaviour of the system is
mathematically chaotic.

\end{document}